\documentclass[preprints,article,accept,moreauthors,pdftex,10pt,a4paper]{mdpi}
\ifdefined\myframe
\renewenvironment{myframe}[2]{\section{#1}\begin{frame}{#2}\vspace{-10pt}}{\end{frame}} 
\else
\newenvironment{myframe}[2]{\section{#1}\begin{frame}{#2}\vspace{-10pt}}{\end{frame}}
\fi

\ifdefined\mybullet
\renewcommand{\mybullet}[1]{\vspace{1mm}\\$\bullet$ {\bf #1}\vspace{1mm}\\}
\else
\newcommand{\mybullet}[1]{\vspace{1mm}\\$\bullet$ {\bf #1}\vspace{1mm}\\}
\fi

\ifdefined\mybulletEQ
\renewcommand{\mybulletEQ}[1]{$\bullet$ {\bf #1}\vspace{1mm}\\}
\else
\newcommand{\mybulletEQ}[1]{$\bullet$ {\bf #1}\vspace{1mm}\\}
\fi

\ifdefined\fracd
\renewcommand{\fracd}[2]{\frac{\displaystyle{#1}}{\displaystyle{#2}}}
\else
\newcommand{\fracd}[2]{\frac{\displaystyle{#1}}{\displaystyle{#2}}}
\fi

\ifdefined\recd
\renewcommand{\recd}[1]{\frac{\displaystyle 1}{\displaystyle{#1}}}
\else
\newcommand{\recd}[1]{\frac{\displaystyle 1}{\displaystyle{#1}}}
\fi

\ifdefined\pdd
\renewcommand{\pdd}[2]{\frac{\displaystyle{\partial{#1}}}{\displaystyle{\partial{#2}}}}
\else
\newcommand{\pdd}[2]{\frac{\displaystyle{\partial{#1}}}{\displaystyle{\partial{#2}}}}
\fi

\ifdefined\biggg
\renewcommand{\biggg}[1]{\scalebox{1.2}{\Bigg{#1}}}
\else
\newcommand{\biggg}[1]{\scalebox{1.2}{\Bigg{#1}}}
\fi

\ifdefined\Biggg
\renewcommand{\Biggg}[1]{\scalebox{1.4}{\Bigg{#1}}}
\else
\newcommand{\Biggg}[1]{\scalebox{1.4}{\Bigg{#1}}}
\fi

\ifdefined\eq
\renewcommand{\eq}[1]{\begin{equation}{#1}\end{equation}}
\else
\newcommand{\eq}[1]{\begin{equation}{#1}\end{equation}}
\fi

\ifdefined\eql
\renewcommand{\eql}[2]{\begin{equation}\label{e:#1}{#2}{\end{equation}}
\else
\newcommand{\eql}[2]{\begin{equation}\label{e:#1}{#2}\end{equation}}
\fi

\ifdefined\ali
\renewcommand{\ali}[1]{\begin{align}{#1}\end{align}}
\else
\newcommand{\ali}[1]{\begin{align}{#1}\end{align}}
\fi

\ifdefined\nl
\renewcommand{\nl}{{\vspace{0.2 cm}\\}}
\else
\newcommand{\nl}{{\vspace{0.2 cm}\\}}
\fi

\ifdefined\Tr
\renewcommand{\Tr}{{\mathrm Tr}}
\else
\newcommand{\Tr}{{\mathrm Tr}}
\fi

\ifdefined\BR
\renewcommand{\BR}{{\mathbb R}}
\else
\newcommand{\BR}{{\mathbb R}}
\fi

\ifdefined\BZ
\renewcommand{\BZ}{{\mathbb Z}}
\else
\newcommand{\BZ}{{\mathbb Z}}
\fi

\ifdefined\BN
\renewcommand{\BN}{{\mathbb N}}
\else
\newcommand{\BN}{{\mathbb N}}
\fi

\ifdefined\BC
\renewcommand{\BC}{{\mathbb C}}
\else
\newcommand{\BC}{{\mathbb C}}
\fi

\ifdefined\Re
\renewcommand{\Re}{\operatorname{Re}}
\else
\newcommand{\Re}{\operatorname{Re}}
\fi

\ifdefined\Im
\renewcommand{\Im}{\operatorname{Im}}
\else
\newcommand{\Im}{\operatorname{Im}}
\fi

\ifdefined\arch
\renewcommand{\arch}{\operatorname{ar\,ch}}
\else
\newcommand{\arch}{\operatorname{ar\,ch}}
\fi

\ifdefined\arsh
\renewcommand{\arsh}{\operatorname{ar\,sh}}
\else
\newcommand{\arsh}{\operatorname{ar\,sh}}
\fi

\ifdefined\arth
\renewcommand{\arth}{\operatorname{ar\,th}}
\else
\newcommand{\arth}{\operatorname{ar\,th}}
\fi

\ifdefined\ch
\renewcommand{\ch}{\operatorname{ch}}
\else
\newcommand{\ch}{\operatorname{ch}}
\fi

\ifdefined\sh
\renewcommand{\sh}{\operatorname{sh}}
\else
\newcommand{\sh}{\operatorname{sh}}
\fi

\ifdefined\th
\renewcommand{\th}{\operatorname{th}}
\else
\newcommand{\th}{\operatorname{th}}
\fi

\ifdefined\Ln
\renewcommand{\Ln}{\operatorname{Ln}}
\else
\newcommand{\Ln}{\operatorname{Ln}}
\fi

\ifdefined\tg
\renewcommand{\tg}{\operatorname{tg}}
\else
\newcommand{\tg}{\operatorname{tg}}
\fi

\ifdefined\ctg
\renewcommand{\ctg}{\operatorname{ctg}}
\else
\newcommand{\ctg}{\operatorname{ctg}}
\fi

\ifdefined\intl
\renewcommand{\intl}{\int\limits}
\else
\newcommand{\intl}{\int\limits}
\fi

\ifdefined\ointl
\renewcommand{\ointl}{\oint\limits}
\else
\newcommand{\ointl}{\oint\limits}
\fi

\ifdefined\integrated
\renewcommand{\integrated}[3]{\left\{{#1}\right\}\left.\vphantom{#1}\right|_{#2}^{#3}}
\else
\newcommand{\integrated}[3]{\left\{{#1}\right\}\left.\vphantom{#1}\right|_{#2}^{#3}}
\fi

\ifdefined\pd
\renewcommand{\pd}[2]{\frac{\partial{#1}}{\partial{#2}}}
\else
\newcommand{\pd}[2]{\frac{\partial{#1}}{\partial{#2}}}
\fi

\ifdefined\rec
\renewcommand{\rec}[1]{\frac{1}{#1}}
\else
\newcommand{\rec}[1]{\frac{1}{#1}}
\fi

\ifdefined\gvec
\renewcommand{\gvec}[1]{\mbox{\boldmath${#1}$}}
\else
\newcommand{\gvec}[1]{\mbox{\boldmath${#1}$}}
\fi

\ifdefined\cvec
\renewcommand{\cvec}[1]{\mbox{\boldmath${#1}$}}
\else
\newcommand{\cvec}[1]{\mbox{\boldmath${#1}$}}
\fi

\ifdefined\td
\renewcommand{\td}[2]{\frac{d{#1}}{d{#2}}}
\else
\newcommand{\td}[2]{\frac{d{#1}}{d{#2}}}
\fi

\ifdefined\md
\renewcommand{\md}[2]{\frac{\mathrm{d}{#1}}{\mathrm{d}{#2}}}
\else
\newcommand{\md}[2]{\frac{\mathrm{d}{#1}}{\mathrm{d}{#2}}}
\fi

\ifdefined\z
\renewcommand{\z}[1]{\left({#1}\right)}
\else
\newcommand{\z}[1]{\left({#1}\right)}
\fi

\ifdefined\ae
\renewcommand{\ae}[1]{\left|{#1}\right|}
\else
\newcommand{\ae}[1]{\left|{#1}\right|}
\fi

\ifdefined\sz
\renewcommand{\sz}[1]{\left[{#1}\right]}
\else
\newcommand{\sz}[1]{\left[{#1}\right]}
\fi

\ifdefined\kz
\renewcommand{\kz}[1]{\left\{{#1}\right\}}
\else
\newcommand{\kz}[1]{\left\{{#1}\right\}}
\fi

\ifdefined\m
\renewcommand{\m}[1]{\mathrm{#1}}
\else
\newcommand{\m}[1]{\mathrm{#1}}
\fi

\ifdefined\c
\renewcommand{\c}[1]{\mathcal{#1}}
\else
\newcommand{\c}[1]{\mathcal{#1}}
\fi

\ifdefined\v
\renewcommand{\v}[1]{\mathbf{#1}}
\else
\newcommand{\v}[1]{\mathbf{#1}}
\fi

\ifdefined\Eq
\renewcommand{\Eq}[1]{Eq.~(\ref{#1})}
\else
\newcommand{\Eq}[1]{Eq.~(\ref{#1})}
\fi

\ifdefined\Eqs
\renewcommand{\Eqs}[2]{Eqs.~(\ref{#1}) and (\ref{#2})}
\else
\newcommand{\Eqs}[2]{Eqs.~(\ref{#1}) and (\ref{#2})}
\fi

\ifdefined\a
\renewcommand{\a}[1]{\aref({#1})}
\else
\newcommand{\a}[1]{\aref({#1})}
\fi

\ifdefined\A
\renewcommand{\A}[1]{\Aref({#1})}
\else
\newcommand{\A}[1]{\Aref({#1})}
\fi

\ifdefined\r
\let\R\r
\renewcommand{\r}[1]{(\ref{#1})}
\else
\newcommand{\r}[1]{(\ref{#1})}
\fi

\ifdefined\comm
\renewcommand{\comm}[2]{\left[{#1},{#2}\right]}
\else
\newcommand{\comm}[2]{\left[{#1},{#2}\right]}
\fi

\ifdefined\Follows
\renewcommand{\Follows}{\qquad\Rightarrow\qquad}
\else
\newcommand{\Follows}{\qquad\Rightarrow\qquad}
\fi

\ifdefined\follows
\renewcommand{\follows}{\quad\Rightarrow\quad}
\else
\newcommand{\follows}{\quad\Rightarrow\quad}
\fi

\ifdefined\followse
\renewcommand{\followse}{\quad\Rightarrow}
\else
\newcommand{\followse}{\quad\Rightarrow}
\fi

\ifdefined\bfollows
\renewcommand{\bfollows}{\Rightarrow\quad}
\else
\newcommand{\bfollows}{\Rightarrow\quad}
\fi

\ifdefined\equivalent
\renewcommand{\equivalent}{\quad\Leftrightarrow\quad}
\else
\newcommand{\equivalent}{\quad\Leftrightarrow\quad}
\fi

\ifdefined\obs
\renewcommand{\obs}[1]{\left\langle{#1}\right\rangle}
\else
\newcommand{\obs}[1]{\left\langle{#1}\right\rangle}
\fi

\ifdefined\ket
\renewcommand{\ket}[1]{\left|{#1}\right\rangle}
\else
\newcommand{\ket}[1]{\left|{#1}\right\rangle}
\fi

\ifdefined\bra
\renewcommand{\bra}[1]{\left\langle{#1}\right|}
\else
\newcommand{\bra}[1]{\left\langle{#1}\right|}
\fi

\ifdefined\braket
\renewcommand{\braket}[2]{\left<#1\vphantom{#2}\right|\left.#2\vphantom{#1}\right>}
\else
\newcommand{\braket}[2]{\left<#1\vphantom{#2}\right|\left.#2\vphantom{#1}\right>}
\fi

\ifdefined\ketbra
\renewcommand{\ketbra}[2]{\left|#1\vphantom{#2}\right>\left<#2\vphantom{#1}\right|}
\else
\newcommand{\ketbra}[2]{\left|#1\vphantom{#2}\right>\left<#2\vphantom{#1}\right|}
\fi

\ifdefined\matrixel
\renewcommand{\matrixel}[3]{\left<#1\vphantom{#2#3}\right|#2\left|#3\vphantom{#1#2}\right>} 
\else
\newcommand{\matrixel}[3]{\left<#1\vphantom{#2#3}\right|#2\left|#3\vphantom{#1#2}\right>} 
\fi

\ifdefined\contravcov
\renewcommand{\contravcov}[3]{{{#1}^{#2}_{}}_{#3}}
\else
\newcommand{\contravcov}[3]{{{#1}^{#2}_{}}_{#3}}
\fi

\ifdefined\covcontrav
\renewcommand{\covcontrav}[3]{{{#1}_{#2}^{}}^{#3}}
\else
\newcommand{\covcontrav}[3]{{{#1}_{#2}^{}}^{#3}}
\fi

\ifdefined\am
\renewcommand{\am}{{\hat{a}^{\vphantom\dagger}}}
\else
\newcommand{\am}{{\hat{a}^{\vphantom\dagger}}}
\fi

\ifdefined\ap
\renewcommand{\ap}{{\hat{a}^\dagger}}
\else
\newcommand{\ap}{{\hat{a}^\dagger}}
\fi

\ifdefined\bm
\renewcommand{\bm}{{\hat{b}^{\vphantom\dagger}}}
\else
\newcommand{\bm}{{\hat{b}^{\vphantom\dagger}}}
\fi

\ifdefined\bp
\renewcommand{\bp}{{\hat{b}^\dagger}}
\else
\newcommand{\bp}{{\hat{b}^\dagger}}
\fi

\ifdefined\arctg
\renewcommand{\arctg}{\mathrm{arc}\,\mathrm{tg}}
\else
\newcommand{\arctg}{\mathrm{arc}\,\mathrm{tg}}
\fi

\firstpage{1} 
\makeatletter 
\setcounter{page}{\@firstpage} 
\makeatother
\pubvolume{xx}
\issuenum{1}
\articlenumber{5}
\pubyear{2018}
\copyrightyear{2018}
\history{Received: date; Accepted: date; Published: date}
\Title{Polarized baryon production in heavy ion collisions: an analytic hydrodynamical study}
\newcommand{\orcidauthorA}{0000-0000-000-000X} 
\Author{B.~Boldizs\'ar $^{1}$\orcidA{}, M.~I.~Nagy $^{1}$ and M.~Csan\'ad $^{1}$}
\AuthorNames{B.~Boldizs\'ar, M.~I.~Nagy, M.~Csan\'ad}
\address{%
$^{1}$ \quad E{\"o}tv{\"o}s Lor{\'a}nd University, H-1117 Budapest, P{\'a}zm{\'a}ny P. s. 1/A, Hungary
}
\corres{Correspondence: nmarci@elte.hu}

\abstract{
We utilize known exact analytic solutions of perfect fluid hydrodynamics to analytically calculate the polarization of baryons produced in heavy ion
collisions. Assuming local thermodynamical equilibrium also for spin degrees of freedom, baryons get a net polarization at their formation (freeze-out).
This polarization depends on the time evolution of the Quark-Gluon Plasma (QGP), which can be described as an almost perfect fluid. By using exact
analytic solutions, we thus can analyze the necessity of rotation (and vorticity) for non-zero net polarization. In this paper we give the first
analytical calculations for the polarization four-vector. We use two hydrodynamical solutions; one is the spherically symmetric Hubble flow (a
somewhat oversimplified model, to demonstrate the methodology). The other solution which we use is a somewhat more involved one that corresponds to a
rotating and accelerating expansion, and is thus well suited to investigate some main features of the time evolution of the QGP created in
peripheral heavy-ion collisions (although there are still many numerous features of a real collision geometry that are beyond the reach of this
simple model). Finally we illustrate and discuss our results on the polarization.
}

\begin{document}
\section{Introduction}

Our aim is to give analytical results for the polarization four-vector of massive spin 1/2 particles produced in heavy-ion collisions,
from hydrodynamical models. The motivation for this work is the recently observed non-vanishing polarization of $\Lambda$ baryons at the STAR
experiment~\cite{STAR:2017ckg,Adam:2018ivw} 
that hints at local thermal equilibrium also for spin degrees of freedom in the Quark Gluon Plasma (QGP) produced in heavy-ion collisions.
The assumption of thermal equilibration for spin is at the core of the current understanding of polarization of particles produced from
a thermal ensemble (such as the QGP), and almost all studies aimed at describing it in terms of collective models utilize the formula
derived from this assumption by Becattini et al.~\cite{Becattini:2013fla}.

Although many numerical hydrodynamical models do indeed predict non-zero polarization of produced spin 1/2 particles~\cite{Csernai:2014nva,
Xie:2016fjj,Karpenko:2016jyx,Xie:2017upb}, a clear connection between the initial state, the final state and the observable polarization
is to be expected from analytical studies, on which topic we do the first calculations here (to our best knowledge).

The observable quantities at the final state of the hydrodynamical evolution can be described by utilizing kinetic theory.
At local thermodynamical equilibrium, for spin 1/2 particles such a description can be based on the the Fermi--Dirac distribution:
\begin{align}
\label{e:fermidirac}
f(x,p) \propto \frac{1}{\exp\z{\dfrac{p_\mu u^\mu (x)}{T(x)} {-} \dfrac{\mu(x)}{T(x)}} {+} 1},
\end{align}
where $p_\mu$ is the four-momentum of the produced particle, and $u^\mu(x)$, $\mu(x)$ and $T(x)$ are the four-velocity, the chemical potential, and
the temperature field of the fluid, respectively.

Assuming local thermal equlibrium for the spin degrees of freedom, for the space-time-- and momentum-dependent polarization four-vector
$\langle S(x,p) \rangle^\mu$ of the produced particles the following formula is given in Ref.~\cite{Becattini:2013fla}:
\begin{align}
\label{e:polarization}
\langle S(x,p) \rangle^\mu = \frac{1}{8m}\big(1 {-} f(x,p)\big) \varepsilon^{\mu\nu\rho\sigma} p_\sigma \partial_\nu \beta_\rho,
\end{align}
where $m$ is the mass of the investigated particle, and the inverse temperature field $\beta^\mu {=} u^\mu /T(x)$ is introduced.
Here $\varepsilon^{\mu\nu\rho\sigma}$ is the totally antisymmetric Levi-Civita-symbol; the $\varepsilon^{0123}=1$ convention is used.
In this paper we use this formula to calculate the polarization four-vector at the freeze-out from analytical relativistic
hydrodynamical solutions.

The general consensus is that the appearance of polarization strongly depends on the rotation of the expanding QGP fireball.
However, the Equation of State (EoS) of the QGP influences the rotation, thus by measuring the polarization, we can get
information about the EoS of the QGP. Analytic hydrodynamic calculations may provide special insight by yielding analytic
formulas for the connections of the aforementioned physical quantities.

We investigate two hydrodynamical solutions: the spherically symmetric Hubble flow~\cite{Csorgo:2003rt,Csorgo:2003ry} and a rotating and accelerating
solution (first reported in Ref.~\cite{Nagy:2009eq}, then in a different context in~\cite{Hatta:2014gqa}). We expect to obtain zero
polarization in the case of the spherical symmetric Hubble-flow as it has no rotation, so the study of this solution can be regarded
as a simple cross-check of our methodology. The second one, however, being a rotating and expanding solution, could be a well usable
model of peripheral heavy-ion collisions, and it is expected that one gets non-zero polarization out of it. Thus this rotating
expanding solution constitutes the core point of the reported work.
 
\section{Basic equations and assumptions}

We use the $c{=}1$ notation. Let us denote the space-time coordinate by $x^\mu{\equiv}(t,\v r)$, and the Minkowskian metric
tensor by $g^{\mu\nu}{=}\m{diag}(1,{-}1,{-}1,{-}1)$. The convention for the Levi-Civita symbol is $\varepsilon^{0123}{=}1$. Greek letters
denote Lorentz indices, Latin letters denote three-vector indices. For repeated Greek indices we use the Einstein summation convention.
We denote the space dimension by $d$; this implies ${g^\mu}_\mu = d{+}1$. In reality, $d=3$, but it is useful to retain the $d$ notation wherever
possible, in order to see if the reason for a specific numeric constant in the formulas is the dimensionality of space.
The four-velocity of the fluid is $u^\mu{=}\gamma(1,\v v)$, where $\gamma{=}\sqrt{1 {-} v^2}$ is the Lorentz factor. The velocity
three-vector is then $\v v{=}u^k/u^0$. With $p^\mu$ we denote the four-momentum of a produced particle; we also use the three-momentum $\v p$, whose
magnitude we simply denote by $p$ (whenever there is no risk of confusion). The energy of the particle is denoted by $E$; the mass shell condition
then reads as $E = \sqrt{p^2{+}m^2}$, with $m$ being the particle mass. 

The usability of hydrodynamics in heavy ion physics phenomenology relies on the assumption of local thermodynamical equilibrium of the matter.
For describing particles with spin 1/2, we use the source function as written up in Eq.~(\ref{e:fermidirac}). Hadronic final state observables
can be then calculated by integrating over the freeze-out hypersurface; e.g. in the case of the invariant momentum distribution, the driving formula is
\begin{align}
\label{e:cooperfrye}
 E \frac{\m dN}{\m d^3 \v p} = \int \m d^3\Sigma_\mu(x) p^\mu f(x,p).
\end{align}
Here $\m d^3 \Sigma_\nu$ is the 3-dimensional vectorial integration measure of the freeze-out hypersurface; the appearance of which is the so-called
Cooper-Frye prescription~\cite{Cooper:1974mv} for calculating the invariant momentum distribution. Of the two solutions (mentioned above) which we
investigate in this work, in the case of the rotating and expanding accelerating solution, we also calculate the invariant momentum distribution,
as this has not been done before.

The formula given in Ref.~\cite{Becattini:2013fla} for the polarization of spin 1/2 particles, as written up in Eq.~\r{e:polarization}, may be utilized
for any given $\beta^\mu{=}u^\mu/T$ field that one gets from a given solution of the hydrodynamical equations.
We are interested in calculating the polarization at the final state of the hydrodynamical evolution, so we must integrate
the $\langle S(x,p)\rangle^\mu$ field over the freeze-out hypersurface. The formula to be analyzed further, that is, that for the observed polarization
$\langle S(p) \rangle^\mu$ of particles with momentum $p$, thus becomes
\begin{align}
\label{e:pol}
\langle S(p) \rangle^\mu = \frac{\int \m d^3 \Sigma_\nu p^\nu f(x,p) \langle S(x,p) \rangle^\mu}{\int \m d^3\Sigma_\nu p^\nu f(x,p)},
\end{align}
as written up e.g. in \cite{Karpenko:2016jyx}.
For being able to perform analytical calculations, we have to make some assumptions. We use saddle point integration, in which one assumes that the
integrand is of the form $f(\v r)g(\v r)$, where $f(\v r)$ is a slowly changing function, while $g(\v r)$ has a unique and sharp maximum; then
the integral can be calculated with a Gaussian approximation as
\begin{align}
\label{e:saddlepoint}
\int \m d^d\v r\,f(\v r) g(\v r) \approx f(\v R_0) g(\v R_0)\sqrt{\frac{(2\pi)^d}{\det\v M }},\qquad
\begin{array}{ll}\textnormal{where}&\quad \v M_{ij} = \partial_i\partial_j g(\v r)\big|_{{\v r} = {\v R_0}},\\
\textnormal{and}&\quad \partial_kg(\v R_0) = 0,
\end{array}
\end{align}
that is, $\v R_0$ is the location of the unique maximum of $g(\v r)$ and $\v M$ is the second derivative matrix.

Another assumption concerns the expression of $\langle S(x,p) \rangle^\mu$, Eq.~\r{e:polarization}: if the exponent in the Fermi--Dirac
distribution is large (i.e. phase space occupancy is small), we can use the Maxwell--Boltzmann distribution instead:
\begin{align}
\label{e:MB}
f(x,p)\ll 1\quad\follows\quad
f(x^\mu ,p^\mu) = \frac{g}{(2\pi\hbar)^d}\exp\big(\frac{\mu(x)}{T(x)} {-} \frac{p_\mu u^\mu}{T(x)} \big).
\end{align}
Here $g$ is the spin-degeneracy factor; for spin 1/2 baryons, $g=2$.

In high energy heavy ion phenomenology (when the collision energy is high enough, say for collisions at RHIC or LHC), the $\mu/T$ factor can
(and usually is) neglected; we use this approximation here\footnote{
The vanishing of $\mu$ can also be interpreted as an absence of a conserved particle number density $n$.
All our conclusions would change only by a proportionality factor if we said $\mu/T = \m{const}$ instead of $\mu/T = 0$; if $\mu\neq0$, we would have had
to introduce $n$. Depending on the equation of state of the matter (one that also contains the conserved particle density $n$), one could write
the $f(x,p)$ function in another form, where the normalization $\int\m dp\,f(x,p) = n(x)$ is evident. For example, if one chooses an ultra-relativistic
ideal gas, with $p = nT$, $\varepsilon = \kappa p$, with $\kappa = d$ as EoS, one has $\frac{g}{(2\pi\hbar)^d}e^{\mu/T} = \frac{n}{4\pi T^3}$.
Indeed, in the solutions discussed below, $\mu/T=$const is satisfied, which means $n\propto T^d$, which is the well-known condition for an adiabatic
expansion.
}. With this we have
\begin{align}
\label{e:MB}
f(x^\mu ,p^\mu) = C_0\exp\big({-}p_\mu \beta^\mu(x)\big),\qquad\textnormal{where}\quad \beta^\mu(x) = \frac{u^\mu(x)}{T(x)},\quad\textnormal{and}\quad
C_0 = \frac{g}{(2\pi\hbar)^d}.
\end{align}
If the Maxwell-Boltzmann approximation is justified, it means that $f(x,p)\ll 1$ indeed, and then also Eqs. \r{e:polarization} and \r{e:pol} become
simpler:
\begin{align}
\label{e:polarizationsimple}
\langle S(x,p) \rangle^\mu = \frac{1}{8m} \varepsilon^{\mu\nu\rho\sigma} p_\sigma \partial_\nu \beta_\rho,
\end{align}
and in the saddle-point approximation, the polarization of particles with momentum $p$ becomes simply
\begin{align}
\label{e:polsimple}
\langle S(p) \rangle^\mu \approx \frac{1}{8m} \varepsilon^{\mu\nu\rho\sigma} p_\sigma \partial_\nu \beta_\rho \Big|_{\v r {=} \v R_0 },
\end{align}
since in the saddle-point approximation, in the numerator of Eq.~\r{e:pol}, $\langle S(x,p) \rangle^\mu$ can be considered the ,,smooth'' function,
and the determinant factors cancel.

\section{Some exact hydrodynamical solutions and polarization}

In this section we first specify and recapitulate the investigated hydrodynamical solutions, then give the analytical formulas for the polarization
four-vector calculated from them. The equations of perfect fluid relativistic hydrodynamics utilized here are 
\begin{align*}
(\varepsilon{+}p)u^\nu\partial_\nu u^\mu &= (g^{\mu\nu}-u^\mu u^\nu)\partial_\nu p &&\textnormal{(Euler equation)},\\
(\varepsilon{+}p)\partial_\mu u^\mu &= -u^\mu\partial_\mu\varepsilon               &&\textnormal{(energy conservation equation)},\\
n\partial_\mu u^\mu &= -u^\mu\partial_\mu n                                        &&\textnormal{(particle number/charge conservation)},
\end{align*}
and we specify the simple $\varepsilon = \kappa p$ equation of state here. (The notations: $\varepsilon$, $p$ and $n$ are the energy density, pressure
and particle number density, respectively.) Concerning the $n$ density: if it is assumed to be non-vanishing, we set the EoS as $p = nT$. However,
the solutions presented below are valid also if $n = 0$ (ie.\@ if $\mu = 0$). So the expressions for $n$ that we recapitulate for the solutions
can be regarded as supplemental to the solutions that work for $\mu = 0$.

We also note that there is recent development on taking the effect that polarization of the constituents of the fluid has on the fluid dynamics
itself~\cite{Florkowski:2018fap}, along with some numerical calculations of how this modified hydrodynamical picture affects final state
polarization~\cite{Florkowski:2019qdp}.
We do not investigate this possibility here; we restrict ourselves to the simple and well-known basic equations witten up above.

\subsection{Hubble flow} 
We do not go into the details about the method to find or verify that the solutions presented below are indeed
solutions of the perfect fluid hydrodynamical equations; we refer back to the original publications of the solutions. 

We investigate the Hubble-like relativistic hydrodynamical solution first fully described in Ref.~\cite{Csorgo:2003rt}. This solution
has the following velocity, particle density and temperature fields:
\begin{align}
\label{e:hubble}
u^\mu & = \frac{x^\mu}{\tau},&
n &= n_0 \left(\frac{\tau_0}{\tau}\right)^d,&
T &= T_0 \left(\frac{\tau_0}{\tau}\right)^{d/\kappa},
\end{align}
where $\tau = \sqrt{t^2-\v r^2}$, and $\kappa$ is the inverse square speed of sound (constant in the case of this exact solution). The $\kappa {=} 3$
case corresponds to ultrarelativistic ideal gas, $\kappa {=} 3/2$ corresponds to a non-relativistic gas; however, this solution is valid for any
arbitrary constant $\kappa$ value\footnote{We note that a more general class of solutions is possible~\cite{Csorgo:2003rt,Csorgo:2003ry,Csanad:2014dpa}
in which the temperature and density fields are supplemented with an arbitrary $\c V$ function of a ,,scaling variable'' $S$:
$$ n = n_0 \left(\frac{\tau_0}{\tau}\right)^d\c V(S),\qquad  T = T_0 \left(\frac{\tau_0}{\tau}\right)^{d/\kappa}\rec{\c V(S)},$$
and the $S$ variable is any function of $S_x$, $S_y$, and $S_z$:
$$
S\equiv S(S_x,S_y,S_z),\quad\textnormal{where}\quad
S_x \equiv \frac{r_x^2}{\dot X_0^2t^2},\quad
S_y \equiv \frac{r_y^2}{\dot X_0^2t^2},\quad
S_z \equiv \frac{r_z^2}{\dot X_0^2t^2},\quad\textnormal{for example:}\quad S =
\frac{r_x^2}{\dot X_0^2t^2}    + \frac{r_y^2}{\dot Y_0^2t^2}    + \frac{r_z^2}{\dot Z_0^2t^2}.$$
Here $\dot X_0$, $\dot Y_0$ and $\dot Z_0$ are arbitrary constants. In the given example, the $S=\m{const}$ surfaces are ellipsoids,
and $\dot X_0$, $\dot Y_0$, $\dot Z_0$ are time derivatives of the principal axes of them}.

To calculate the polarization four-vector, as of now we investigate the simplest case, the spherical symmetric expansion.
For the freeze-out hypersurface the $\tau{=}\tau_0{=}\m{const.}$ hypersurface is chosen (which in the case of the investigated solution equals the constant
temperature freeze-out hypersurface), and a given point of this hypersurface can be parametrized simply by the $\v r$ coordinate three-vector,
and the time coordinate on the hypersurface is $t(\v r){\equiv}\sqrt{\tau_0^2{+}\v r^2}$. The integration measure
and the resulting expression for the Cooper--Frye formula can then be written as
\begin{align}
\label{e:mb}
\m d^3\Sigma_\mu = \rec{t(\v r)}\begin{pmatrix} t(\v r) \\ \v r \end{pmatrix}\m d^3\v r
\quad\follows\quad
 E\frac{\m dN}{\m d^3\v p} =
C_0\int \m d^3\v r\frac{ E t(\v r){-}\v p\v r}{t(\v r)}\exp\z{-\frac{ E t(\v r){-}\v p{\v r}}{T_0}}.
\end{align}
As we are discussing massive particles, this integral always exists. The $T_0$ constant (an arbitrary parameter of the solution) can be taken simply
as the temperature at freeze-out; we did so. 

The position of the saddle-point ($\v R_0$) as well as the second derivative matrix $M_{kl}$ is calculated as
\begin{align}
\partial_k\frac{ E t{-}\v p{\v r}}{T_0}\Big|_{\v r{=}\v R_0}\stackrel{!}{=}0 \follows\v R_0=\frac{\tau_0}{m}\v p.
\qquad\qquad
M_{kl}\equiv -\partial_k\partial_l\frac{ E t-\v p{\v r}}{T_0}\Big|_{\v r=\v R_0} = 
\frac{m}{T_0\tau_0}\z{\delta_{kl}-\frac{p_k p_l}{ E^2}}.
\end{align}
With this we can get an approximation for the invariant single-particle momentum distribution:
\begin{align}
\det \v M=\frac{m^2}{ E^2}\z{\frac{m}{T_0\tau_0}}^3\quad\follows\quad
 E\frac{\m dN}{\m d^3\v p} = \frac{n_0}{4}\sqrt{\frac{\pi \tau_0^3}{mT_0^3}}\exp\z{-\frac{\tau_0 m}{T_0}}.
\end{align}
The formula is independent of momentum. This was expected because this hydrodynamical solution (in the $\c V(S) {=} 1$ case) is boost invariant.

To use (\ref{e:polsimple}) to determine the polarization four-vector in the hydrodynamical solution of the Hubble-flow, first we give the expression
for the $\partial_\nu \beta_\rho$ derivative:
\begin{align}
\partial_\nu \beta_\rho = \partial_\nu \z{\frac{r_\rho}{\sqrt{\tau_0^2 {+} r^2}T_0}}=
\frac{g_{\nu\rho}}{\sqrt{\tau_0^2 {+} r^2}T_0}+\frac{r_\nu r_\rho}{(\tau_0^2 {+} r^2)^{3/2}T_0}.
\end{align}
Then for the time component we get:
\begin{align}
\langle S(p) \rangle^0 =\frac{1}{8mT_0} \varepsilon^{0ikl} p_l\partial_i \beta_k\bigg|_{\v r =\v R_0} =
\rec{8mT_0} \varepsilon_{ikl}p_l \z{\frac{g_{ik}}{\sqrt{\tau_0^2 {+} r^2}T_0}+\frac{r_{i}r_{k}}{(\tau_0^2 {+} r^2)^{3/2}T_0}}\Bigg|_{\v r {=} \v R_0}=0,
\end{align}
as $\varepsilon^{0ikl}$ is antisymmetric whereas $g_{ik}$ and $r_i r_k$ are symmetric to the change in the $i\leftrightarrow k$ indices.

Similarly for the spatial coordinates:
\begin{align}
\langle S(p)\rangle^i = \rec{8mT_0}
\bigg(-\varepsilon_{ikl}p_l\partial_0\beta_k+\varepsilon_{ikl}p_l\partial_k\beta_0-\varepsilon_{ikl}p_0\partial_k\beta_l\bigg)\Bigg|_{\v r=\v R_0}=0.
\end{align}
In conclusion, the polarization four-vector in the spherical symmetric Hubble-flow is
\begin{align}
\langle S(p)\rangle^\mu = \begin{pmatrix} 0\\ \v 0\\ \end{pmatrix},
\end{align}
which is consistent with our expectations.

\subsection{Rotating and accelerating expanding solution}\label{sec:rotpolarization}

Another hydrodynamical solution of particular interest to us is a rotating and accelerating expanding solution, first written up in Ref.~\cite{Nagy:2009eq}.
This solution has the following velocity, temperature and particle density profiles:
\begin{align}
\label{e:rot}
\v v &= \frac{2t\v r {+} \tau_0^2\gvec\Omega{\times}\v r}{t^2{+}r^2{+}\rho_0^2},&
T&= \frac{T_0\tau_0^2}{\sqrt{(t^2{-}r^2{+}\rho_0^2)^2{+}4\rho_0^2 r^2{-}\tau_0^4(\gvec\Omega{\times}\v r})^2},&
n &= n_0\z{\frac{T}{T_0}}^3,
\end{align}
where $\rho_0$ and $\tau_0$ are arbitrary parameters and $\gvec\Omega$ is an arbitrary angular velocity three-vector that indicates the axis and magnitude
of rotation. The $\rho_0$ parameter tells about the initial spatial extent of the expanding matter, however, the $\tau_0$ parameter is just there for the
sake of consistency of physical units; in this way, the unit of $\gvec\Omega$ is $c$/fm, as it should be for an angular velocity-like
quantity\footnote{Here we changed the notation of Ref.~\cite{Nagy:2009eq}. The rather unfortunate $\v B$ notation used there is now written
as $\tau_0^2\gvec\Omega$.}, and $T_0$ is a temperature constant.
In the case of $\gvec\Omega {=} 0$, we recover an acceleratingly expanding but non-rotating spherically symmetric solution.

It is convenient to write up this solution with the following notation:
\begin{align}
\label{e:solclass}
&\frac{u^\mu}{T} \equiv \beta^\mu  = a^\mu {+} F^{\mu\nu}x_\nu {+} (x^\nu b_\nu) x^\mu {-} \frac{x^\nu x_\nu}{2}b^\mu,
\\\textnormal{with}\quad&
a^\mu {=} \frac{\rho_0^2}{2T_0\tau_0^2}\begin{pmatrix}1\\\v 0\end{pmatrix},\qquad b^\mu {=} \rec{T_0\tau_0^2}\begin{pmatrix}1\\\v 0\end{pmatrix}, \qquad
F_{0k} {=} F_{k0} {=} F_{00} {=} 0, \qquad F_{kl} {=} \varepsilon_{klm} \frac{\Omega_m}{2T_0}.
\end{align}
To calculate final state observables, we choose the constant proper time ($\tau_0=$const) hypersurface here as well.
The solution itself allows for a re-scaling of the arbitrary constants in the formulas; just as in the previous case, here too
we can treat the $T_0$ quantity as the temperature at freeze-out (at the $\v r = 0$ center of the expanding matter).
We use the notation introduced in Eq.~(\ref{e:mb}) for the Maxwell--Boltzmann distribution. To derive the saddle point
for the calculation of the polarization four-vector, we shall use the expression of the invariant momentum spectrum:
\begin{align}
\label{e:rotcooperfrye}
 E\frac{\m dN}{\m d^3\v p} = C_0\int\m d^3\v r\z{ E{-}\frac{\v p\v r}{\sqrt{\tau_0^2{+}r^2}}}
\exp\kz{-\frac{E(2r^2 {+} \tau_0^2{+}\rho_0^2){-}2\sqrt{\tau_0^2 {+} r^2}\v p\v r{-}\tau_0^2\v r(\v p{\times}\gvec\Omega)}{T_0\tau_0^2}}.
\end{align}
This integral always exists (in the case of massive particles).
In order to utilize the saddle-point integration method, we determine the position of the saddle point ($\v R_0$) and the second derivative matrix
at the saddle point:
\begin{align}
\label{e:R0eq:rotaccel}
&\textnormal{for $\v R_0$}:\qquad
\nabla\Big\{ {-} \rec{T_0\tau_0^2}\Big( E(2r^2 {+} \tau_0^2 {+} \rho_0^2) - 2\sqrt{\tau_0^2 {+} r^2}\v r\v p-\tau_0^2\v r(\v p{\times}\gvec\Omega)\Big)\Big\}\Big|_{\v r = \v R_0}
\stackrel{!}{=}0,\\
&M_{kl} = \partial_k\partial_l
\Big\{\frac{1}{T_0\tau_0^2}\Big( E(2r^2 {+} \tau_0^2 {+} \rho_0^2) - 2\sqrt{\tau_0^2 {+} r^2}\v r\v p-\tau_0^2\v r(\v p{\times}\gvec\Omega)\Big)\Big\}\Big|_{\v r = \v R_0}.
\label{e:Mrotaccel}
\end{align}
We leave the detailed calculations to Appendix~\ref{sec:rotappendix}; the results are the following. The $\v R_0$ saddle point (for a given $\v p$ momentum)
is in the plane spanned by the $\v p$ and $\v p{\times}\gvec\Omega$ vectors. In the following we use the $\hat{\v p}\equiv \v p/p$ notation for the unit
vector pointing in the direction of $\v p$. For the saddle point we get
\begin{align}\label{e:R0rotaccel}
\v R_0 = \frac{\tau_0}{2p}\sqrt{\frac{E{-}m}{2m}}\sqrt{\tau_0^2(\hat{\v p}{\times}\gvec\Omega)^2(E{-}m)^2+4p^2}\cdot\hat{\v p} +
\tau_0^2\frac{E{-}m}{2p}\cdot\hat{\v p}{\times}\gvec\Omega.
\end{align}
Concerning the second derivative matrix, we need it only for the calculation of the invariant momentum distribution, where its determinant is invoked.
It turns out that this quantity is
\begin{align}
\label{e:detMrotaccel}
\operatorname{det}M_{kl} = \frac{32m^2}{T_0^3\tau_0^6}(E{+}m)p.
\end{align}
Using this result, we get the invariant single-particle momentum distribution\footnote{This has not yet been calculated for this hydrodynamical solution.}
as
\begin{align}
 E\frac{\m dN}{\m d^3\v p} \propto \sqrt{\frac{\pi^3T_0^3\tau_0^3}{32p(m{+}E)}}\exp\z{-\frac{E_{\textnormal{eff}}}{T_0}},
\quad\textnormal{with}\quad
E_{\textnormal{eff}}=m{+}\frac{\rho_0^2 E}{\tau_0^2}{+}\frac{\tau_0^2}{4}(\gvec\Omega^2{-}(\hat{\v p}\gvec\Omega)^2)\z{ E{-}m}.
\end{align}
Equivalently, by defining a ``local slope'' $T_{\textnormal{eff}}$, the result can be expressed as
\begin{align}
 E\frac{\m dN}{\m d^3\v p} \propto \sqrt{\frac{\pi^3T_0^3\tau_0^3}{32p(m{+}E)}}\exp\z{-\frac{E}{T_{\textnormal{eff}}}},
\quad\textnormal{with}\quad
T_{\textnormal{eff}}=\frac{T_0}{\frac{m}{E}{+}\frac{\rho_0^2}{\tau_0^2}{+}\frac{\tau_0^2}{4}(\gvec\Omega^2{-}(\hat{\v p}\gvec\Omega)^2)\z{1{-}\frac{m}{E}}}.
\end{align}
Proceeding to the polarization of the produced baryons, we calculate the derivative of the inverse temperature field for this solution from the form
given in Eq.~(\ref{e:solclass}), then substitute it into the expression of the polarization, Eq.~(\ref{e:polsimple}). The result is
\begin{align}
\partial_\nu \beta_\rho =
  F_{\rho\nu} {+} x^\alpha b_\alpha g_{\nu\rho} {+} x_\rho b_\nu {-} x_\nu b_\rho \quad\follows\quad
\langle S(p)\rangle^\mu =
  \rec{8m}\varepsilon^{\mu\nu\rho\sigma}p_\sigma
  \Big(F_{\rho\nu} {+} x_\rho b_\nu {-} x_\nu b_\rho \Big)\Big|_{\v r {=} \v R_0}.
\end{align}
(The second term was cancelled owing to the symmetry of $g_{\nu\rho}$ and the antisymmetry of $\varepsilon^{\mu\nu\rho\sigma}$, and
$x^\mu$ is understood as the four-coordinate of the freeze-out hypersurface whose three-coordinate is the $\v r=\v R_0$ three-vector).
Remembering the expression of the introduced $F^{\mu\nu}$ tensor and $b^\mu$ vector from Eq.~(\ref{e:solclass}), in particular that $F^{0k}=0$, and $b^k=0$,
we get the following expressions for the the time-like and space-like components: 
\begin{align}
\langle S(p) \rangle^0 &= {-} \rec{8m}\varepsilon^{0klm} p_m (F_{kl} {+} x_l b_k {-} x_k b_l)\Big|_{\v r {=} \v R_0}
= -\rec{16m} \varepsilon_{klm}\varepsilon_{klq}p_m\frac{\Omega_q}{T_0} = \rec{8m}\frac{\v p\gvec\Omega}{T_0},\\
\langle S(p)\rangle^k &= \rec{8m}\bigg(\varepsilon^{k0lr}p_r(F_{l0} {+} x_l b_0 {-} x_0 b_l){+} \varepsilon^{kl0r}p_r(F_{0l} {+} x_0 b_l {-} x_l b_0){+}
\varepsilon^{klr0}p_0(F_{rl} {+} x_r b_l {-} x_l b_r)\bigg)\Big|_{\v r{=}\v R_0} =\nonumber\\&=
-\rec{8m}\bigg(2b_0\varepsilon_{klm}x_lp_m {+} E\varepsilon_{klm}\varepsilon_{mlq}\frac{\Omega_q}{2T_0}\bigg)\Big|_{\v r {=} \v R_0}
= \rec{8mT_0}\Big(E\gvec\Omega - \frac{2}{\tau_0^2}\v R_0{\times}\v p\Big)_k =\nonumber\\&= \frac{m\Omega_k {+} (E{-}m)\hat p_l\Omega_l\hat p_k}{8mT_0}.
\end{align}
Summarizing this result, the polarization four-vector for the investigated rotating and accelerating expanding solution is the following:
\begin{align}
\label{forgopol}
\langle S(p)\rangle^\mu =\frac{1}{8mT_0} \begin{pmatrix} \v p\gvec\Omega \\ m\gvec\Omega +\frac{E{-}m}{p^2}(\gvec\Omega\v p)\v p \end{pmatrix}.
\end{align}
In the case of $\gvec\Omega = 0$, there is no rotation, and we get $\langle S(p)\rangle^\mu {=} 0$. In this model thus polarization is very
transparently connected to the presence of rotation.

It is useful to transform the polarization four-vector into the rest frame of the particle. The result is\footnote{
The Lorentz matrix performing this boost transformation is the following (in usual 1+3 dimensional block matrix notation):
$${\Lambda^\mu}_\nu =\begin{pmatrix} \cosh\chi & -\hat p_l\sinh\chi \\
-\hat p_k\sinh\chi & \delta_{kl} {+} (\cosh\chi{-}1)\hat p_k\hat p_l \end{pmatrix}
= \rec m\begin{pmatrix} E & -p_l \\ -p_k & m\delta_{kl} +\frac{E{-}m}{p^2}p_kp_l \end{pmatrix},$$
where $E$ and $p$ could be parametrized with the velocity parameter $\chi$ as $E = m\cosh\chi$ and $p = m\sinh\chi$, respectively. It indeed
can be checked that this matrix takes the $(E,\v p)$ four-momentum vector into $(m,\v 0)$, as it should.}, with (r.f. standing for ``rest frame''):
\begin{align}\label{e:polSrf}
\langle S(p) \rangle^\mu_{\textnormal{r.f.}} = \begin{pmatrix} 0 \\ \v S_{\textnormal{r.f.}} \end{pmatrix},
\quad\textnormal{where}\quad \v S_\textnormal{r.f.} = \rec{8T_0}\gvec\Omega. 
\end{align}
We can also compute the helicity of the produced spin 1/2 particles in this solution from this formula (the $\v S$ polarization vector is taken in the
laboratory frame):
\begin{align}
\label{e:helicity}
H := \hat{\v p}\v S = \frac{E}{8mT_0}\hat{\v p}\gvec\Omega.
\end{align}

\section{Illustration and discussion}

\begin{figure}[h!]
\centering
\includegraphics[width=0.9\columnwidth]{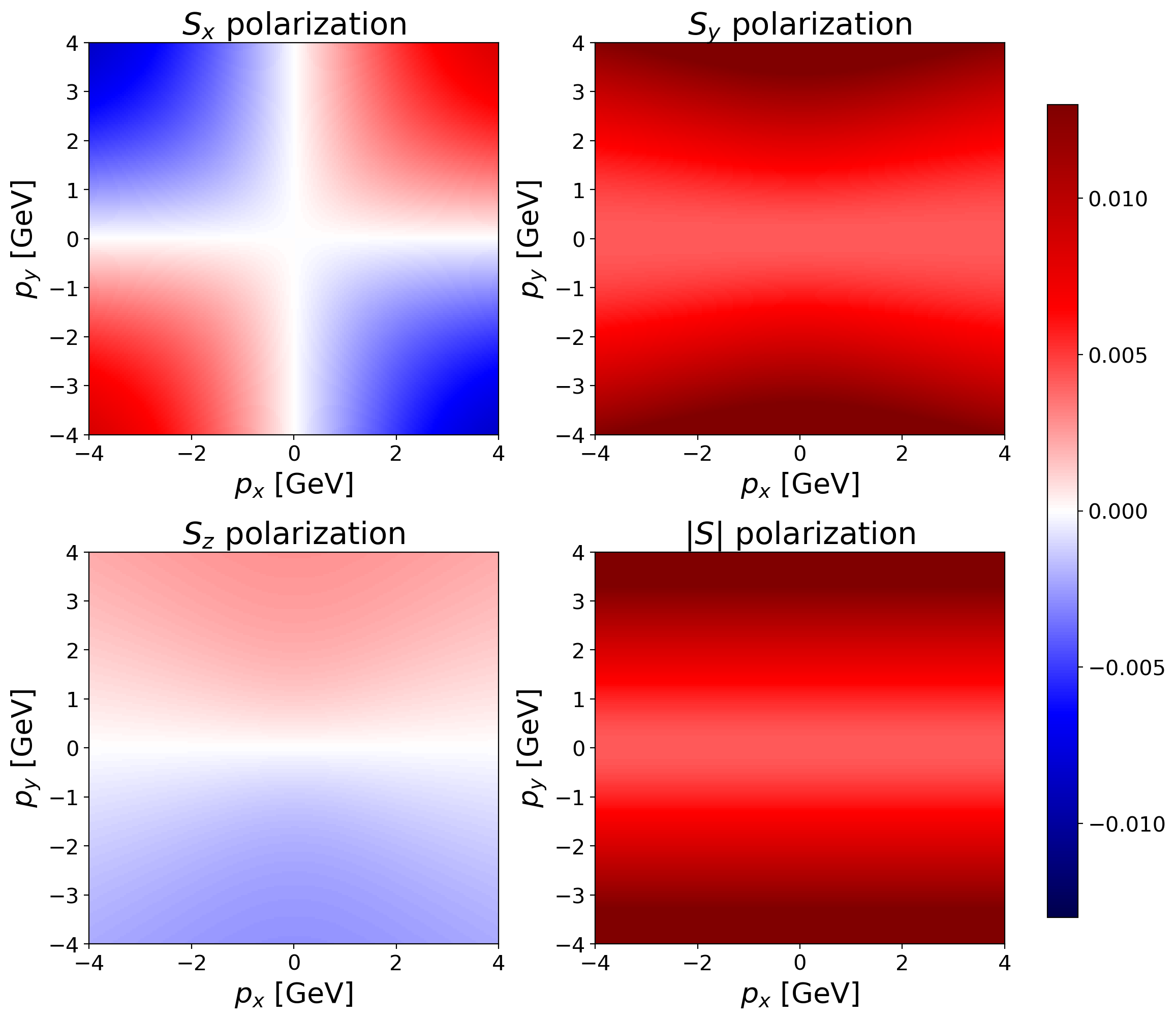}
\caption{The components of the polarization four-vector in the rotating and accelerating expanding solution with respect to the momentum. Plots were
made with the mass of the $\Lambda$ baryon ($m_\Lambda = 1115$ MeV$/c^2$), and with $|\gvec\Omega| = 0.1\,c/\m{fm}$.}
\label{f:pol}
\end{figure}

\begin{figure}[h!]
\centering
\includegraphics[width=0.5\columnwidth]{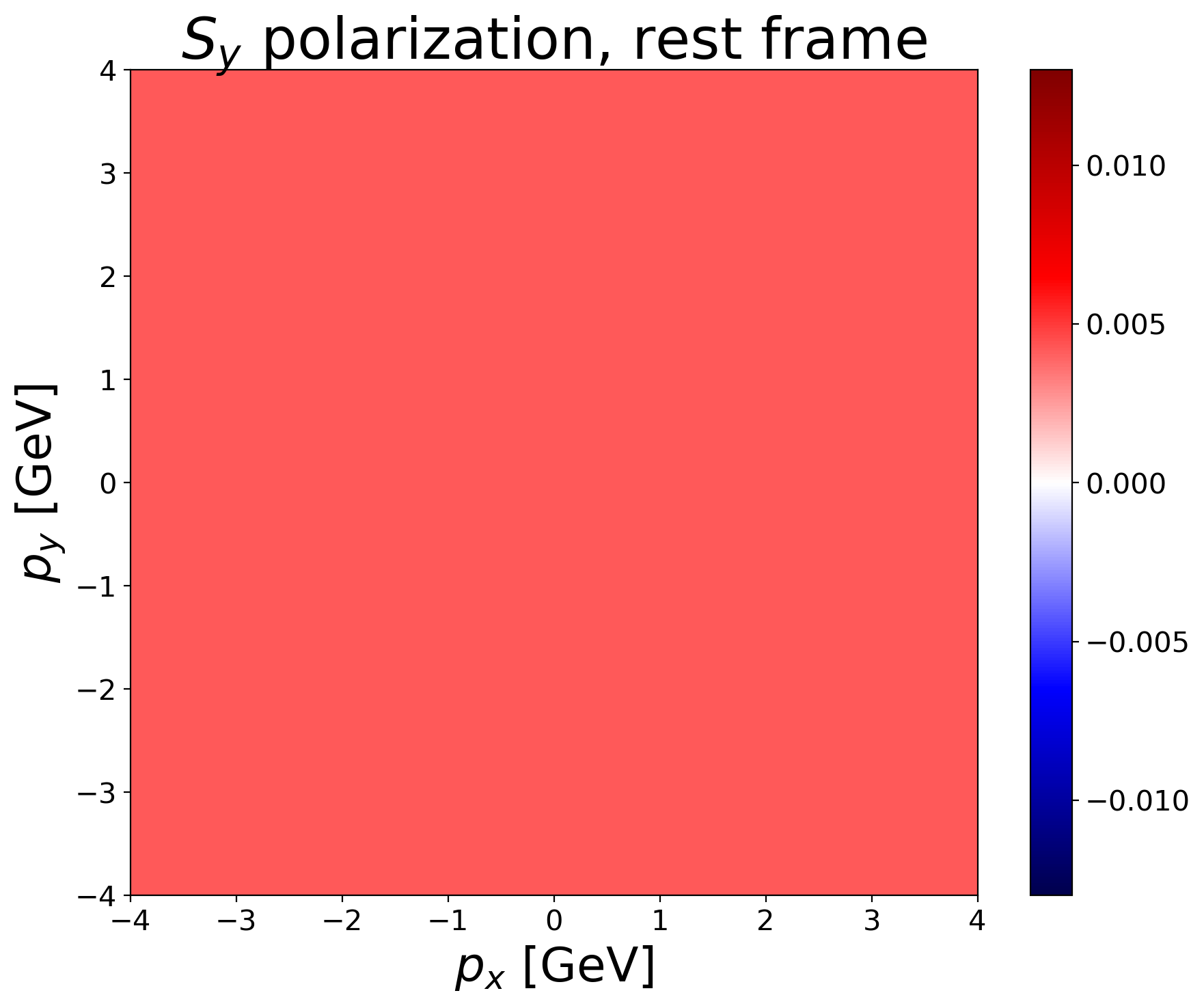}
\caption{
The only non-vanishing component of the polarization vector in the rest frame of the baryon is $S_y$ in the investigated simple solution; in this case
its value is uniquely determined by the magnitude of the $\gvec\Omega$ vector. More involved types of analytic solutions would yield some dependence on the
momentum components $p_x$ and $p_y$. For the plotted value of $S_y$ (a constant, as seen in the plot) the same input parameters were used as above:
$m_\Lambda = 1115$ MeV$/c^2$, and $|\gvec\Omega| = 0.1\,c/\m{fm}$.}
\label{f:pol2}
\end{figure}

\begin{figure}[h!]
\centering
\includegraphics[width=0.55\columnwidth]{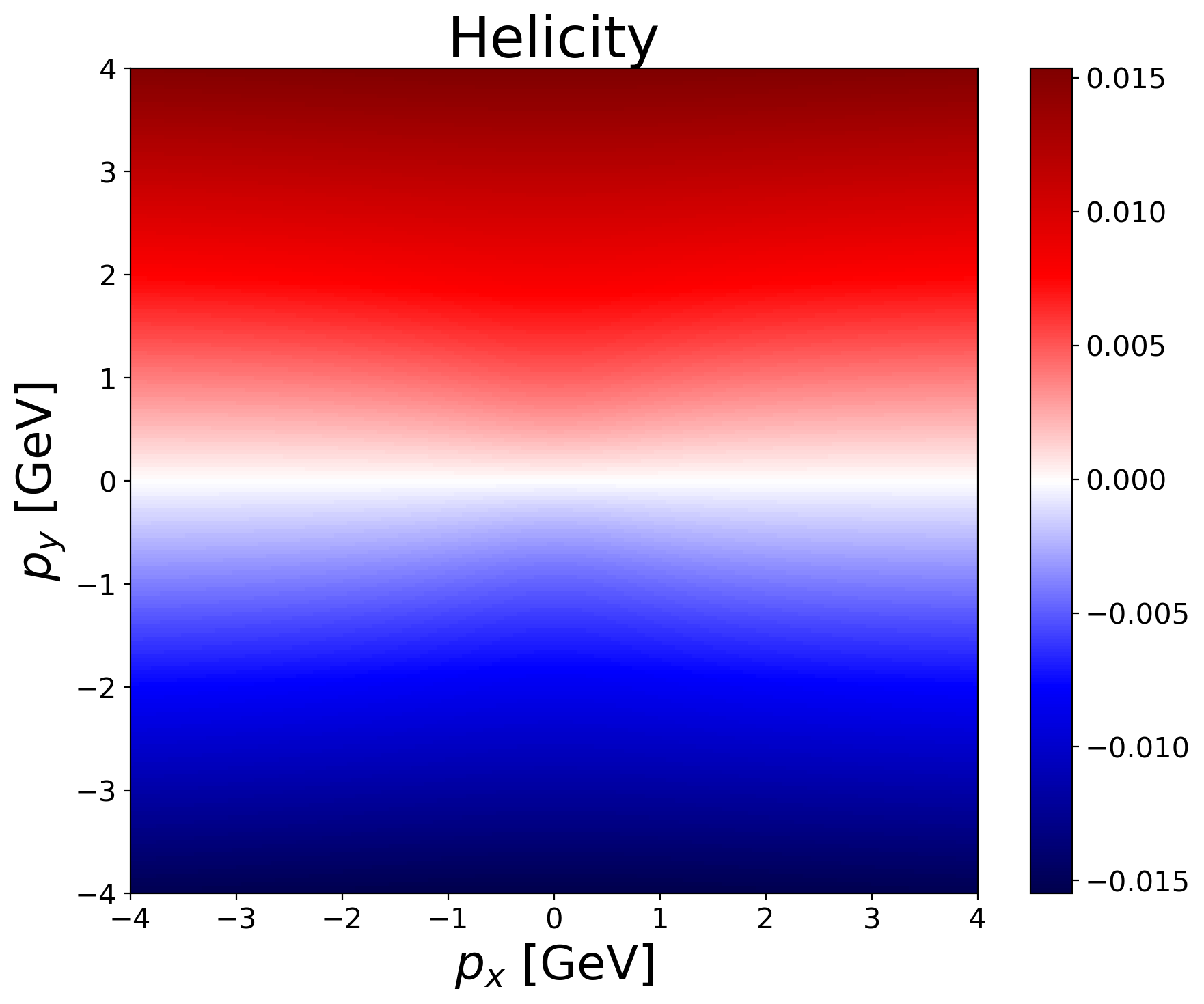}
\caption{Helicity of the produced baryons calculated in the rotating and accelerating expanding solution. Parameter values as above: $m_\Lambda = 1115$ MeV$/c^2$, $|\gvec\Omega| = 0.1\,c/\m{fm}$.}
\label{f:helicity}
\end{figure}

In this section, we would like to illustrate our simple analytic results for the polarization vector. We use the same type of plots that was used to
visualize some existing numerical simulations (e.g. those presented in Ref.~\cite{Karpenko:2016jyx}). We plot the components of the polarization
vector with respect to the momentum components in the transverse plane (that is, w.r.t. $p_x$ and $p_y$). On Fig.~\ref{f:pol} we plot the polarization
vector in the laboratory frame. For the sake of plotting, the mass of the $\Lambda$ baryon ($m_\Lambda c^2 = 1115$ MeV) was chosen. For the sake of this
illustration, we chose a moderate value for the magnitude of the $\gvec\Omega$ vector as $|\gvec\Omega|=0.1\,c/\m{fm}$.
In our case, as a special coincidence owing purely to the specific algebraic form of the presented analytic solution, it turned out that the polarization
in the rest frame of the produced baryons is independent of momentum $\v p$; see \Eq{e:polSrf}. This coincidence is expected to be relieved in the case
of more involved (complicated) solutions (that are left for future investigations). Fig.~\ref{f:pol2} nevertheless shows the value of the $S_y$ component
in the baryon rest frame.
The helicity of the produced baryons (being proportional to the $\v p\v S$ scalar product), however, {\it does} depend on the momentum, even in the case
of our very simple solution. We plot it on Fig.~\ref{f:helicity}; with the same parameter values as in the foregoing two plots.

\section{Summary and outlook}

In this paper we gave the first analytical formulas for the polarization of baryons produced from a thermal ensemble corresponding to rotating and expanding
exact hydrodynamical solutions. These arise as descriptions of the final state of non-central high energy heavy-ion collisions. We investigated two exact
relativistic hydrodynamical solutions. One was the spherically symmetric Hubble flow (an overly simplistic one, the study of which can be regarded as
a check of the methodology), in which the polarization turns out to be exactly zero (as it is naturally expected from symmetry considerations).
The other solution we investigated is a one describing rotating and accelerating expansion. In this case we got the first ever analytical formulas that
connect dynamical quantities of the expansion (i.e.\@ magnitude of rotation, acceleration, etc) with the observable final state polarization of
spin 1/2 particles (baryons), which turns out to be non-zero in this case.

Our results are simple and straightforward. Nevertheless, many more solutions (more involved ones) as well as more complicated final state
parametrizations can be investigated in the future. The calculations presented here yield the first results in terms of exact formulas for the
polarization; more refined future studies are needed to disentangle the effects that rotation, acceleration and temperature gradient have
on the observable final state polarization of baryons produced in heavy-ion collisions. Such studies have the potential of a better understanding
of what phenomenological implications can polarization measurements (such as recently done by the STAR experiment~\cite{STAR:2017ckg})
have on the properties (such as the equation of state) of the strongly coupled Quark Gluon Plasma produced in heavy ion collisions.

\section*{Acknowledgements}
\funding{Our research has been partially supported by the Hungarian NKIFH grants No. FK-123842 and FK-123959, the Hungarian EFOP 3.6.1-16-2016-00001
project. M. Csan\'ad and M.~Nagy was supported by the J\'anos Bolyai Research Scholarship of the Hungarian Academy of Sciences and the
\'UNKP New National Excellence Program of the Hungarian Ministry of Human Capacities.}

\appendix
\subsection{Additional calculations}
\label{sec:rotappendix}
Here we discuss some additional calculations used in Section \ref{sec:rotpolarization} pertaining to the case of rotating and accelerating solution.

For a given momentum $\v p$, the position of the saddle point $\v R_0$ (to be applied in the approximate calculation of the momentum spectrum and
the polarization) was written up in \Eq{e:R0rotaccel}; we provide some additional details of the derivation of that formula here.
The defining equation was \Eq{e:R0eq:rotaccel}, of which the following equation for $\v R_0$ is obtained:
\begin{align}
4E\v R_0 - 2\sqrt{\tau_0^2{+}R_0^2}\v p - \frac{2(\v p\v R_0)}{\sqrt{\tau_0^2{+}R_0^2}}\v R_0 - \tau_0^2(\v p{\times}\gvec\Omega) = 0,
\end{align}
where $R_0^2{\equiv}\v R_0\v R_0$. From this equation one readily sees that $\v R_0$ must be a linear combination of $\v p$ and
the $\v p{\times}\gvec\Omega$ vector. We substitute this assumption into the equation above. We note that $\v p$ and $\v p{\times}\gvec\Omega$
are orthogonal to each other, which leads to some intermediate simplifications, as well as enables us to rearrange the obtained condition into
the following form:
\begin{align*}
\v R_0:=\alpha\v p + \beta\tau_0^2\v p{\times}\gvec\Omega \follows 2\kz{\z{2 E {-} \frac{\alpha p^2}{A}}\alpha {-} A}\v p =
\tau_0^2\kz{1{-}2\beta\z{2E{-}\frac{\alpha p^2}{A}}}(\v p{\times}\gvec\Omega).
\end{align*}
where we temporarily introduced the $A{\equiv}\sqrt{\tau_0^2 {+} \alpha^2p^2 + \beta^2\tau_0^4(p^2\Omega^2{-}(\v p\gvec\Omega)^2)}$ notation.
Because of the orthogonality of $\v p$ and $\v p{\times}\gvec\Omega$, both sides here have to vanish identically, from which we get
\begin{align}
A&=\alpha\z{2 E {-} \frac{\alpha p^2}{A}}, & 4 E {-} \frac{2\alpha p^2}{A}&=\frac{1}{\beta}.
\end{align}
One divides these equations to obtain a simple relation, the substituting back one gets a quadratic equation for $\beta$, the solution of which is
\begin{align}
\frac{\alpha}{\beta}=2A\Follows  4 E {-} 4\beta p^2 = \rec\beta \Follows \beta=\frac{ E}{2p^2}\pm\sqrt{\frac{ E^2}{4p^2}-\frac{p^2}{4p^2}}
= \frac{ E{\pm} m}{2p^2},
\end{align}
where we used the $E^2{=}p^2{+}m^2$ relation. To find $\alpha$ we substitute this back into the expression of $A$:
\begin{align*}
\alpha {=} 2\beta A\follows
\alpha^2 {=} 4\beta^2\kz{\tau_0^2{+}\alpha^2p^2{+}\beta^2\tau_0^4(p^2\Omega^2{-}(\v p\gvec\Omega)^2)} \follows
\alpha {=} 2\beta\tau_0\sqrt{\frac{1{+}\beta^2\tau_0^2(p^2\Omega^2{-}(\v p\gvec\Omega)^2)}{1{-}4p^2\beta^2}}.
\end{align*}
Using the above expression of $\beta$ (with the yet undetermined sign) we get $1{-}4p^2\beta^2 = -\frac{2m}{p^2}(m{\pm}E)$, and see that the expression
for $\alpha$ will be valid only in the case when $1{-}4\beta^2p^2>0$, thus conclude that the bottom sign is the proper choice. We thus arrive at
the following expressions:
\begin{align}
\beta&=\frac{ E {-} m}{2p^2},&
\alpha&= 2\beta\tau_0\sqrt{\frac{1{+}\beta^2\tau_0^2(p^2\Omega^2{-}(\v p\gvec\Omega)^2)}{1{-}4p^2\beta^2}}
= \frac{\tau_0}{2}\sqrt{\frac{E{-}m}{2m}}\sqrt{\tau_0^2(\hat{\v p}{\times}\gvec\Omega)^2(E{-}m)^2+4p^2}.  
\end{align}
From these formulas the expression of $\v R_0$ shown in \Eq{e:R0rotaccel} readily follows.
The other ingredient in the saddle-point integration necessary for getting the momentum spectrum is the determinant of the second derivative matrix of
the source function. Here we outline the main steps of the derivation of \Eq{e:detMrotaccel}. From \Eq{e:Mrotaccel} the second derivative matrix itself
turns out to be
\begin{align}
M_{kl}=\rec{T_0\tau_0^2}\kz{\z{4E{-}\frac{2(\v p\v r)}{A}}\delta_{kl}{-}\frac2A(p_k r_l {+} r_k p_l){+}2(\v p\v r)\frac{r_kr_l}{A^3}}\bigg|_{\v r{=}\v R_0},
\end{align}
where we use the notation $A$ as above. We should use the expression of $\v R_0$ as calculated above.

The determinant of this $\v M$ matrix is the product of its eigenvalues. In our case the particular spatial directions are:
$\v p$, $\v p{\times}\gvec\Omega$, and the vector orthogonal to both these, that is, $\v p{\times}(\v p{\times}\gvec\Omega)$.
One recognizes that the vector $\v p{\times}(\v p{\times}\gvec\Omega)$ is an eigenvector of the $\v M$ second derivative matrix:
\begin{align}
\v M\big(\v p{\times}(\v p{\times}\gvec\Omega)\big) = \dots
= \rec{\beta}\v p{\times}(\v p{\times}\gvec\Omega).
\end{align}
The corresponding eigenvalue is thus $1/\beta$. Owing to the symmetric nature of $\v M$, the other two eigenvectors must be in the orthogonal
complementer subspace of this vector, so they are linear combinations of $\v p$ and $\v p{\times}\gvec\Omega$. Let us thus look for these eigenvectors in
the form $\v a {=} \mu\v p {+} \nu\v r$, with yet to be determined $\mu$ and $\nu$ coefficients. Substituting this expression, we get
\begin{align}
\v M\,\v a = \lambda\v a\follows
\z{4 E {-} \frac{2(\v p\v R_0)}{A}}\v a {-} \frac{2}{A}\bigg(\v R_0(\v a\v p) {+} \v p(\v a\v R_0) \bigg) {+} 2(\v p\v R_0)\frac{\v R_0(\v a\v R_0)}{A^3}
= \lambda \v a,
\end{align}
where $\lambda$ is the eigenvalue (the values of which we are looking for). 
By substituting the assumed form of $\v a$ and inferring the components of this equation in the $\v p$ and $\v p{\times}\gvec\Omega$ directions, we
get the following equation for the $\mu$ and $\nu$ coefficients:
\begin{align}
\frac2{A^3}\begin{pmatrix} 2EA^3{-}2A^2\v p\v R_0 & {-}A^2R_0^2 \\ {-}A^2p^2{+}(\v p\v R_0)^2 & 2EA^3{-}2A^2\v p\v R_0{+}R_0^2\v p\v R_0 \end{pmatrix}
\begin{pmatrix} \mu \\ \nu  \end{pmatrix} = \lambda \begin{pmatrix} \mu \\ \nu \end{pmatrix}.
\end{align}
We immediately infer the product of the two $\lambda_{1,2}$ eigenvalues as the determinant of this $2{\times}2$ matrix. Taking the third eigenvalue
(calculated above) into account, after some simplifications, we indeed get the following expression for the determinant of the $\v M$ matrix (the
expression we used in \Eq{e:detMrotaccel}):
\begin{align}
\det\v M=\z{\rec{T_0\tau_0^2}}^3\, 32pm^2(m{+}\sqrt{p^2{+}m^2}).
\end{align}

\end{document}